\documentstyle[epsf,prb,floats,twocolumn,aps]{revtex}

\begin{document}

\baselineskip 11.5 pt

\title{Non-linear Microwave Surface Impedance of Epitaxial\\
HTS Thin Films in Low DC Magnetic Fields}

\author{A.P.Kharel$^1$, K.H.Soon$^1$, J.R.Powell$^1$,
A. Porch$^1$, M.J.Lancaster$^1$, A. V. Velichko$^{1,2}$ and
R.G.Humphreys$^3$
\vspace*{0.7 true cm}
\\ {\small $^1$School  of Electronics \& Electrical Engineering,
University of Birmingham, UK}
\\ {\small $^2$Institute of Radiophysics \& Electronics of NAS, Ukraine}
\\ {\small $^3$DERA, Malvern WR14 3PS, UK}}

\maketitle

\abstract{%
     We have carried out non-linear microwave (8 GHz) surface impedance
measurements of three YBaCuO thin films in dc magnetic fields $H_{dc}$
(parallel to c axis) up to 12 mT using a coplanar resonator technique.
In zero dc field the three films, deposited by the same method, show a
spread of low-power residual surface resistance, $R_{res}$ and
penetration depth, $\lambda$~($T=15$~K) within a factor of 1.9.
However, they exhibit dramatically different microwave field, $H_{r\!f}$
dependences of the surface resistance, $R_s$, but universal
$X_s(H_{r\!f})$ dependence.  Application of a dc field was found to affect
not only absolute values of $R_s$ and $X_s$, but the functional
dependences $R_s(H_{r\!f})$ and $X_s(H_{r\!f})$ as well. For some of the
samples the dc field was found to decrease $R_s$ below its  zero-field
low-power value.}

\section{Introduction}
Understanding the mechanisms of the non-linearity of high-$T_c$
superconductors (HTS) at microwave frequencies is very important from the
point of view of application of the materials in both passive and active
microwave devices~\cite{Heinrev}. Recently, unusual features such as
decrease of the surface resistance $R_s$ and reactance $X_s$ of HTS thin
films with microwave field $H_{r\!f}$ have been
reported~\cite{Hein97,Khar98}.  Similar observations were made in weak
($\leq 20$~mT) static fields $H_{dc}$~\cite{Hein97,Choudh97}, which have
shown that a small dc magnetic field can cause a decrease of $R_s$ and
$X_s$ in both the linear and nonlinear regimes.

In the present paper we report measurements of the microwave field
dependences of $R_s$ and $X_s$ of high-quality epitaxial YBaCuO thin
films in zero and finite ($\leq 12$~mT) applied dc magnetic fields.
All the samples have rather different functional form of
$R_s(H_{r\!f})$, but $X_s(H_{r\!f})$ is universal and nearly
temperature-independent.  At the same time, $H_{dc}$ applied parallel to
the c-axis of the films has a qualitatively similar effect on both
$R_s(H_{r\!f})$ and $X_s(H_{r\!f})$, giving evidence of non-monotonic
behavior of $R_s$ and $X_s$ as a function of $H_{dc}$ both in the linear
and nonlinear regimes. An even more striking feature is that for some of
the samples the dc field can decrease $R_s$ below its low-power zero-field
value, thereby offering a possible way of reducing the microwave losses of
HTS thin films.

\section{Experimental Results}
\label{exp}
The films are deposited by e-beam co-evaporation onto polished
(001)-oriented  MgO single crystal $10\times10$~mm$^2$ substrates. The
films are 350 nm thick. The c-axis misalignment of the films is typically
less than 1$\%$,  and the $dc$ critical current density $J_c$ at 77~K is
around $2\cdot10^6$ A/cm$^2$~\cite{Chew}. The films were
patterned into linear coplanar transmission line resonators with
resonance frequency of $\sim 8$~GHz using the technique described
in~\cite{Porch}. The nonlinear measurements were performed using
a vector network analyzer with a microwave amplifier providing CW output
power up to 0.3~W. The low-power values of $R_s$ and $\lambda$ at 15~K are
60, 35, 50~$\mu\Omega$  and 260, 210, 135~nm for samples TF1, TF2 and TF3,
respectively.

\begin{figure}[t]
\def\epsfsize#1#2{0.45#1}
\centerline{\epsfbox{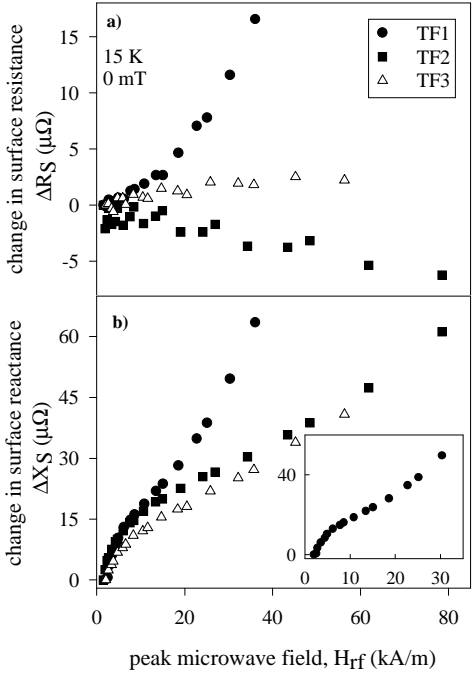}}
\vspace{-1.6 true cm}
\caption{Change~in~(a)~the~surface~resistan\-ce $\Delta R_s
=R_s(H_{r\!f})-R_s(0)$ and (b) the surface reactance
$\Delta X_s=X_s(H_{r\!f})-X_s(0)$ as a function of peak microwave magnetic
field $H_{r\!f}$ for three samples (specified in the figure) at $T=15$~K,
$H_{dc}=0$. The inset shows data for sample TF1 on an expanded scale.}
\label{fig1}
\end{figure}

Changes in $R_s$ and $X_s$ with $H_{r\!f}$, $\Delta
R_s=R_s(H_{r\!f})-R_s(0)$ and $\Delta X_s=X_s(H_{r\!f})-X_s(0)$, are
plotted in Fig.~\ref{fig1} for all three samples. It is seen that the
$H_{r\!f}$-dependence of $\Delta R_s$ is rather different for different
samples, whereas $\Delta X_s(H_{r\!f})$ is universal. For sample TF1,
$\Delta R_s\sim H_{r\!f}^2$ from the
lowest $H_{r\!f}$. For sample TF2, a decrease in $R_s$ is observed
with increased $H_{r\!f}$, and the absolute value of $R_s$ falls below the
corresponding low-power value. Finally, for sample TF3, $\Delta R_s$ is
rather independent of $H_{r\!f}$ up to sufficiently high fields
($\sim$60~kA/m), after which a skewing of the resonance curve is observed.
The surface reactance, $X_s$, for all three samples is a sublinear
function of $H_{r\!f}$ ($\sim H_{r\!f}^n$, $n<1$) at low powers, then has
a kink, followed by a superlinear functional dependence ($\sim
H_{r\!f}^n$, $n>1$).

\begin{figure}[t]
\def\epsfsize#1#2{0.5#1}
\centerline{{\hspace{0.6 true cm}\epsfbox{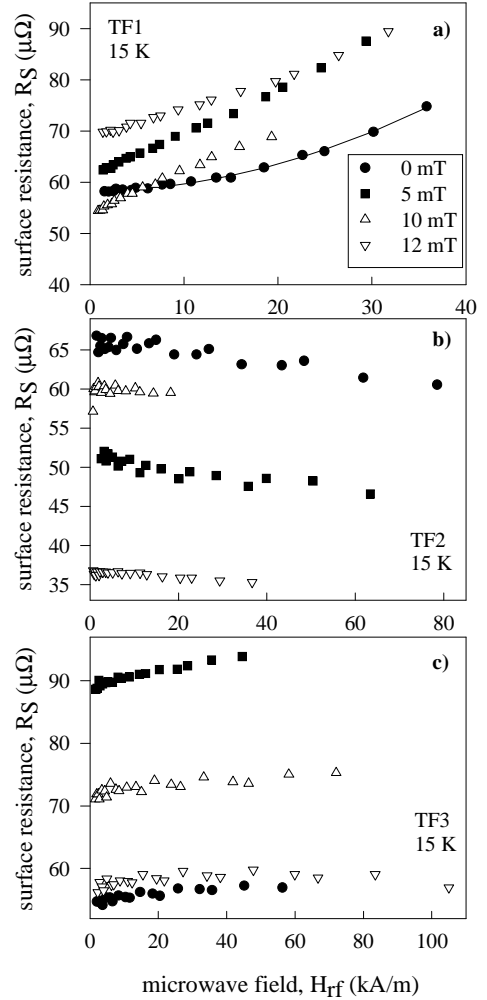}}}
\caption{Surface resistance $R_s$ as a function of $H_{r\!f}$
at different dc magnetic fields 0, 5, 10 and 12~mT and $T=15$~K,
for three samples (a,b,c). The solid curve in (a) shows a fit
$\sim H_{r\!f}^2$ to the data taken at 0~mT.}
\label{fig2}
\end{figure}

\begin{figure}[t]
\def\epsfsize#1#2{0.5#1}
\centerline{\epsfbox{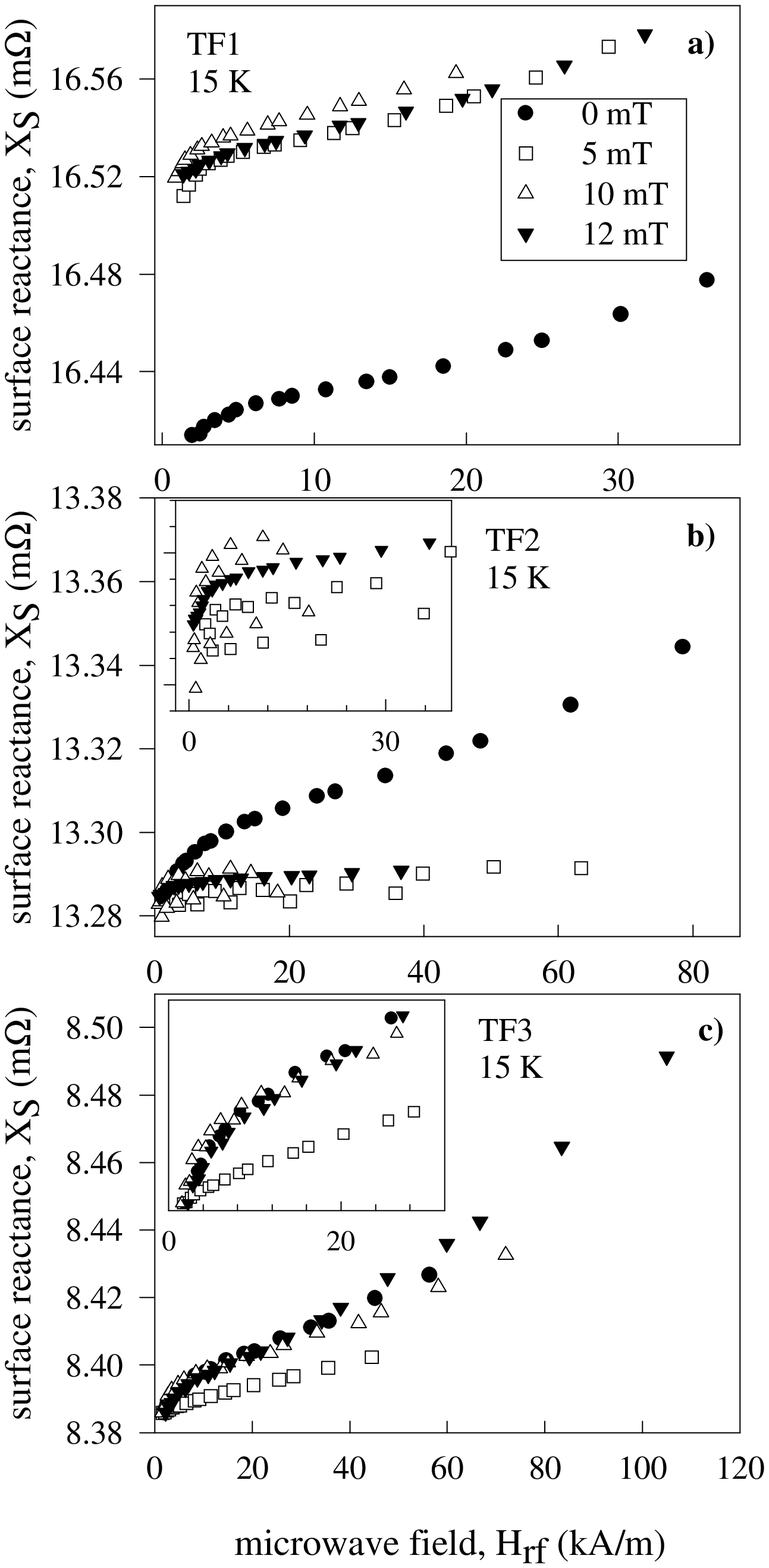}}
\caption{Surface reactance $X_s$ as a function of $H_{r\!f}$
at different dc magnetic fields 0, 5, 10 and 12~mT and $T=15$~K,
for three samples (a,b,c). The insets in (b) and (c) show the low-power
data on an expanded scale.}
\vspace{-0.05 true cm}
\label{fig3}
\end{figure}

The effect of dc magnetic fields ($\leq 12$~mT) on the microwave power
dependence of $R_s$ and $X_s$ for all the samples is illustrated in
Fig.~\ref{fig2} and Fig.~\ref{fig3}. The common feature for all three
samples is that the dependences of $R_s(H_{r\!f})$ and $X_s(H_{r\!f})$
upon $H_{dc}$ are non-monotonic. For samples TF1 and
TF2 (for particular $H_{r\!f}$-range and $H_{dc}$-values), the static
field leads to a decrease in $R_s$ compared to the low-power zero-field
value. This means that both dc and rf fields can cause a {\it reduction of
the microwave losses\/} in YBaCuO (see Fig.~\ref{fig1}a and
Fig.~\ref{fig2}a,b). A possible mechanism of such a behavior is
discussed later.

One can see that for sample TF1, a dc field of a certain strength (10~mT)
can cause a decrease in $R_s$, whereas $X_s$ is always enhanced by a dc
field.  Similarly, for sample TF3, the behavior of $R_s(H_{r\!f})$ and
$X_s(H_{r\!f})$ in $H_{dc}$ is also uncorrelated. However, for TF1
we observe a reduction of $R_s$ without an accompanying decrease in $X_s$,
whereas for TF3 the effect is opposite; for particular values of $H_{dc}$
(5, 10~mT) the in-field ($H_{dc}\neq 0$) value of $X_s(H_{r\!f})$ is lower
than the corresponding value for $H_{dc}=0$ (Fig.~\ref{fig2}c), while the
in-field value of $R_s(H_{r\!f})$ is always higher than corresponding
zero-field value (Fig.~\ref{fig2}c).  Here, the most pronounced decrease
in $X_s$ for TF3 is observed at $H_{dc}=5$~mT.  Finally, for sample TF2
there is a well pronounced correlated behavior of $R_s(H_{r\!f})$ and
$X_s(H_{r\!f})$ in a dc field; $H_{dc}$ of any value from 5 to 12~mT
decreases both $R_s$ and $X_s$ (see Fig.~\ref{fig2}b and
Fig.~\ref{fig3}b).
\section{Discussion and Conclusions}

\begin{figure}[t]
\def\epsfsize#1#2{0.5#1}
\centerline{\epsfbox{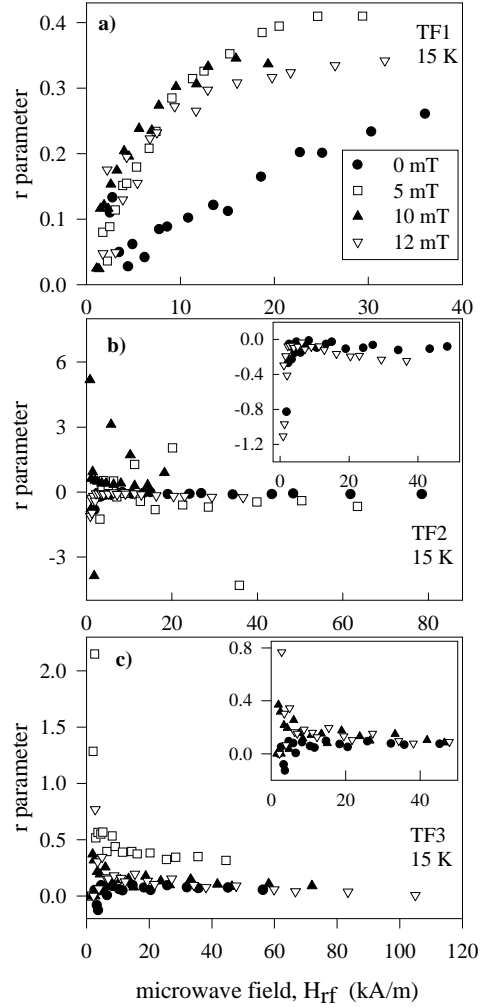}}
\caption{The impedance plane parameter $r=\Delta R_s/\Delta X_s$
as a function of $H_{r\!f}$ at different dc magnetic fields 0, 5, 10 and
12~mT and $T=15$~K, for three samples (a,b,c). The insets in (b) and (c)
show some of the data on an expanded scale.}
\label{fig4}
\end{figure}

A powerful approach in distinguishing between various non-linear mechanisms
is a parametric representation of the data in terms of the $r$ parameter,
where $r=\Delta R_s/\Delta X_s$~\cite{Halb97,Golos95}. In Fig.~\ref{fig4}
we plotted the $H_{r\!f}$-dependence of the $r$ parameter for all three
samples in different dc magnetic fields from 0 to 12~mT. One can see that
for sample TF1, all the in-field $r(H_{r\!f})$ curves almost collapse over
the entire range of $H_{r\!f}$, whereas the zero-field $r(H_{r\!f})$ data
are clearly different from the in-field ones. This is especially
noticeable at low $H_{r\!f}$ (between 3-7~kA/m), where the $r$ values
differ by up to a factor of 10 between the zero-field and in-field
$r(H_{r\!f})$ dependences. At the same time, at the lowest $H_{r\!f}$
(2-3~kA/m), the zero-field $r$-values match very well with the in-field
ones (see Fig.~\ref{fig4}a). Therefore, the low-power nonlinearity for
sample TF1 appears to have the same origin for zero-field and in-field
regimes, whereas the high-power range mechanisms are likely to be
different.

For sample TF2 at $H_{dc}=5$~mT and 10~mT, $r(H_{r\!f})$ is rather noisy,
which clearly correlates with the noisy dependence of $X_s(H_{r\!f})$ for
this sample at the relevant dc fields (see inset in Fig.~\ref{fig3}b). The
$r$ parameter oscillates between $-$4 and 6 with  an average values close
to 0.3--0.4 and 0.2--0.3 for 5 and 10~mT, respectively.  For zero field
and 12 mT, $r(H_{r\!f})$ are quite consistent, starting to increase from
large negative values $\sim -1$ at low powers, and saturating to the level
of $-$0.2 to $-$0.1 at higher $H_{r\!f}$.

Finally, for sample TF3 at zero dc field, $r$ increases with $H_{r\!f}$ at
low powers, whereas the 10 and 12~mT $r$-values decrease, but all three
curves level off for $H_{r\!f}\geq 10$~kA/m to a value of $\sim 0.1$.
However, $r(H_{r\!f})$ at 12~mT appears to tend to small negative values
at high $H_{r\!f}$, consistent with the decrease in $R_s(H_{r\!f})$ at
12~mT in the relevant $H_{r\!f}$ range (Fig.~\ref{fig2}c). Standing apart
from other dependences is $r(H_{r\!f})$ at 5~mT, which shows very high
values ($\sim 2$) at low $H_{r\!f}$, saturating at a level of $\sim 0.4$
at higher $H_{r\!f}$. Note that this value is about a factor of 4 larger
than the saturation level for other curves. This seems to imply that
$H_{dc}=5$~mT causes a switching of the mechanism of the nonlinearity in
the film, as compared to the mechanism at other fields, including
zero-field results.

Recently Ma et al.~\cite{Ma} have found that YBaCuO thin films
deposited by the same method exhibit correlation of $R_s(H_{r\!f})$
with the values of low-power residual $\lambda_{res}$ and the normal-fluid
conductivity $\sigma_n$. At the same time, they failed to note
any correlation between the power dependence and $R_{res}$.
A similar conclusion can be drawn from our results (see Fig.~\ref{fig1}a).
One can see that $R_s$ is almost independent of $H_{r\!f}$ for sample TF3,
which has the lowest $\lambda(15~K)=135$~nm, whereas sample TF1 with the
largest $\lambda(15~K)=260$~nm exhibits the strongest
$H_{r\!f}$-dependence.  On the other hand, there is no strict correlation
between $R_s(H_{r\!f})$ and low-power $R_s$ (see Sec.~\ref{exp} and
Fig.~\ref{fig1}a), which is also consistent with the results of Ma et al.
However, the strongest power dependence, $R_s\sim H_{r\!f}^2$, is observed
for sample TF1 with both the highest $R_{res}$ and $\lambda_{res}$, in
agreement with recent results on YBaCuO thin films with different
low-power characteristics~\cite{Velich}.

  There are two further distinctive features for samples TF1, when
compared with the two other samples. The functional form of
$R_s(H_{r\!f})$ is noticeably changed by a dc field ($R_s\sim H_{r\!f}^n$,
where $n=$2, 1.12, 0.8 and 1.24 for 0, 5, 10 and 12~mT, respectively),
whereas $X_s(H_{r\!f})$ is not affected by $H_{dc}$. In addition, for
TF1, a dc magnetic field changes not only the power dependence of $R_s$,
but the absolute value of the low-power $X_s$ (see Fig.~\ref{fig2}a),
while for TF2 and TF3 no such effect is observed (Fig.~\ref{fig2}b,c). The
effect of a dc field on $R_s(H_{r\!f})$ is also seen for sample TF3 at
$H_{dc}=5$~mT which, as will be argued later, may
switch the mechanism of nonlinearity for this sample.

Recently Habib et al.~\cite{Habib} have found that for a
stripline resonator with a weak link in the middle, $R_s(H_{r\!f})$ is
strongly affected by the junction, whereas $X_s(H_{r\!f})$
was found to be insensitive to the presence of the weak link.
Based on this finding, we can suggest that the difference between
$R_s(H_{r\!f})$ for our samples may originate from different
microstructure (type, dimension and number of defects) of the samples,
whereas the similar form of $X_s(H_{r\!f})$ appears to reflect the
intrinsic behavior of each film, mostly exhibited by grains. This
assumption is further supported by the strong effect of a small dc field
on $R_s(H_{r\!f})$ for samples TF1 (Fig.~\ref{fig2}a) and TF3 (at 5~mT,
Fig.~\ref{fig2}c), whereas the functional form of $X_s(H_{r\!f})$ is
unchanged by $H_{dc}$.

\subsection{Analysis of Possible Mechanisms}

As we have shown earlier~\cite{Khar98}, such uncorrelated behavior of
$R_s(H_{r\!f})$ and $X_s(H_{r\!f})$, as we observed for our samples
(Fig.~\ref{fig1}), cannot be explained by any of the known theoretical
models, including  Josephson vortices~\cite{Halb97} (where $r_{JF}<1$,
$\Delta R_s$,$\Delta X_s\sim H_{r\!f}^n$, $0.5<n<2$), heating of weak
links~\cite{Golos95} ($r_{HE}<1$, $\Delta R_s,\Delta X_s\sim H^2$) and the
RSJ model~\cite{Halb97} ($r_{RSJ}<1$, $\Delta R_s$ increasing in a
stepwise manner and $\Delta X_s$ oscillating with $H_{r\!f}$), intrinsic
pair breaking or uniform heating~\cite{Golos95} (for both mechanisms
$r<10^{-2}$, and $\Delta R_s,\Delta X_s\sim H^2$).  We can also rule out
the mechanism of the superconductivity stimulation by microwave
radiation~\cite{Eliash}, recently claimed by us~\cite{Khar98} and
Choudhury et al.~\cite{Choudh97}. In this mechanism, the dc magnetic field
decreases the order parameter, increasing both $R_s$ and $X_s$, which we
do not observe for any of our samples. Moreover, we see that even in the
low-power regime $H_{dc}$ can cause reduction of both $\Delta R_s$ and
$\Delta X_s$, which is not explained by the above model at all.

The most plausible mechanism responsible for the decrease in $R_s$ and
$X_s$ with both $H_{r\!f}$ and $H_{dc}$ fields seems to be field-induced
alignment of the spins of magnetic impurities, which are likely to be
present in most HTS (particularly in YBaCuO)~\cite{Kres1}. This
mechanism was recently claimed by Hein et al.~\cite{Hein97} to explain
their results on non-monotonic behavior of $R_s$ and $X_s$ in
$H_{dc}$ and $H_{r\!f}$ for YBaCuO thin films. However, because our
non-linear results for $\Delta R_s$ and $\Delta X_s$ are not correlated,
and moreover, exhibit different $R_s(H_{r\!f})$ dependences, we suppose
that other strong nonlinear mechanism(s) may interfere with the
spin-alignment mechanism.  We suggest that this mechanism might be Cooper
pair breaking at low powers, and nucleation and motion of
rf-vortices~\cite{Halb97} at higher powers.  Heating effects at high
powers may also play an important role.  However, additional
investigations are necessary to answer this question unambiguously.

In conclusion, we have presented here the results on non-monotonic
microwave power dependence of $R_s$ and $X_s$ in both zero and weak ($\leq
12$~mT) dc magnetic field for very high-quality epitaxial YBaCuO thin
films. Since this unusual behavior has come into being only
owing to a significant progress in the thin films fabrication for the past
few years, we conclude that the features observed by us seem to originate
from the intrinsic properties of superconductors. However, different
functional form of $R_s(H_{r\!f})$ for different samples and universal
$X_s(H_{r\!f})$ behavior seem to imply that the microstructure still
plays a significant role in the macroscopic properties of the samples.
In addition, the observed decreases in $R_s$ and $X_s$ below their
zero-field low-power values means that there is still room for
improvement of the microwave properties of the thin films. This
can be realized upon adequate understanding of the mechanisms
responsible for the unusual behavior observed, and can lead to
improved characteristics of HTS-based microwave devices.

\end{document}